\documentclass[conference]{IEEEtran}
\IEEEoverridecommandlockouts
\usepackage{cite}
\usepackage{amsmath,amssymb,amsfonts}
\usepackage{algorithmic}
\usepackage{graphicx}
\usepackage{textcomp}
\usepackage{xcolor}
\def\BibTeX{{\rm B\kern-.05em{\sc i\kern-.025em b}\kern-.08em
    T\kern-.1667em\lower.7ex\hbox{E}\kern-.125emX}}

\begin{document}
\title{Capacity Enhancement of n-GHZ State Super-dense Coding Channels by Purification and Quantum Neural Network}

\author{{Rong~Zhang, Xiaoguang~Chen, Yaoyao~Wang and 
			Bin~Lu
		}\\
		\IEEEauthorblockA{
			\textit{Department of Communications Science and Engineering}\\
			\textit{School of Information Science and Technology, Fudan University}\\
               \textit{Shanghai, 200433, China}\\
			\textit{(e-mails:  Xiaoguangchen@fudan.edu.cn)}
		}
}

\maketitle

\begin{abstract}
A super-dense coding protocol based on the n-GHZ state is proposed to enable the two communicating parties to choose the number of transmitted code words according to their demand and to adapt the quantum super-dense coding protocol to multiple transmitted code word scenarios. A method is proposed that combines entanglement purification and Quantum Neural Network (QNN) to improve the channel capacity of super-dense coding. By simulating a realistic quantum communication noise environment in the Cirq platform, the effect of purification and QNN on the enhancement of fidelity and channel capacity in super-dense coding communication scenarios with different dimensions under unitary and non-unitary noise conditions is analyzed. The experimental results show that the channel capacity of super-dense coding is improved in different degrees when purification and QNN are applied separately, and the combination of purification and QNN has a superimposed effect on the channel capacity enhancement of super-dense coding, and the enhancement effect is more significant in different dimensions.
\end{abstract}


\section{Introduction}
Quantum communication is a communication method based on quantum mechanics, which makes use of the special properties of quantum states, such as quantum superposition states, quantum entanglement states, the uncertainty principle and unclonability\cite{vathsan2015introduction}\cite{9797967}, to ensure the security and reliability of communication. Compared with traditional classical communication, quantum communication adopts the special properties of the quantum state, which means the information in the communication process cannot be eavesdropped or tampered with, thus ensuring the security of communication. Secondly, quantum communication takes advantage of the redundancy of quantum states, i.e. multiple quantum states can transmit the same piece of information at the same time, thus improving the reliability of communication.

Quantum dense coding was first proposed by Bennett and Wiesner \cite{bennett1992communication} in 1992 and successfully verified by Mattle and his team \cite{mattle1996dense} in physical experiments in 1996 and is now a widely used quantum communication protocol for QKD\cite{long2002theoretically}. Quantum dense coding based on the Bell state can transmit 2 bits of classical information while transmitting 1 qubit, which in a sense greatly increases the channel capacity of the communication channel. The process of quantum dense coding involves the process of entanglement distribution, and the current methods to improve the efficiency of quantum entanglement distribution mainly include quantum entanglement concentration\cite{bennett1996concentrating}\cite{lo2001concentrating}, quantum error correction coding, and entanglement purification\cite{bennett1996purification}. Among them, entanglement purification improves the fidelity of one pair of entangled states by discarding and retaining another pair of entangled states and improves the entanglement of shared quantum states after entanglement distribution, which is the main method studied in this paper.

Kerstin and his team proposed in 2020 a quantum neural network (QNN) based on the Verdon training method\cite{tacchino2019artificial}, which creates a quantum feed-forward neural network by quantum simulation of classical neurons. This quantum neural network is trained to obtain a unitary operator, which can be used as a component of a quantum circuit to deal with the effects of specific noise.

This paper is organized as follows. In Section \ref{Section2}, the theoretical basics of this chapter will be introduced. In Section \ref{Section3}, we propose a super-dense coding protocol based on n-GHZ state. In Section \ref{Section4} the algorithm of capacity enhancement of n-GHZ state super-dense coding channels based on purification and QNN will be proposed, and simulation and result analysis. In Section \ref{Section5}, we will reach a conclusion.

\section{Preliminary}\label{Section2}
\subsection{Quantum Gate}
Quantum gates are a fundamental element of quantum computing and their role in quantum circuits is somewhat similar to that of logic gates in classical computers.

In this paper, we use $\sigma_x$, $\sigma_y$, $\sigma_z$ and CNOT gates. The matrix expressions are as follows
\begin{equation}
    \sigma_x=\begin{bmatrix}
 0 & 1\\
 1 & 0
\end{bmatrix},\sigma_y=\begin{bmatrix}
 0 & -i\\
 i & 0
\end{bmatrix},\sigma_z=\begin{bmatrix}
 1 & 0\\
 0 & -1
\end{bmatrix}
\end{equation}

\begin{equation}
    CNOT=\begin{bmatrix}
  1&  0&  0&0 \\
  0&  1&  0&0 \\
  0&  0&  0&1 \\
  0&  0&  1&0
\end{bmatrix}
\end{equation}

\begin{figure}[htbp]
\centerline{\includegraphics[width=0.3\linewidth]{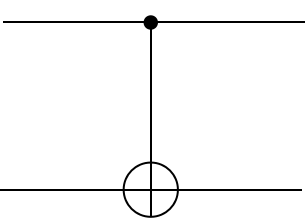}}
\caption{The quantum circuit diagram symbol of the CNOT gate.}
\label{figx1}
\end{figure}

The quantum circuit diagram symbol for the CNOT gate is shown in Fig.~\ref{figx1}

\subsection{GHZ state}
A GHZ (Greenberger-Horne-Zeilinger) state is a quantum state in which three qubits are fully entangled, or an n-GHZ state if it contains $n(n\ge 3)$ qubits.
The GHZ state is represented as follows
\begin{equation}
    \begin{split}
\left |\Psi_1  \right \rangle=\frac{1}{\sqrt{2}}  (\left |000\right \rangle+\left |111\right \rangle),\left |\Psi_2  \right \rangle=\frac{1}{\sqrt{2}}  (\left |000\right \rangle-\left |111\right \rangle)\\
\left |\Psi_3  \right \rangle=\frac{1}{\sqrt{2}}  (\left |001\right \rangle+\left |110\right \rangle),\left |\Psi_4  \right \rangle=\frac{1}{\sqrt{2}}  (\left |001\right \rangle-\left |110\right \rangle)\\
\left |\Psi_5  \right \rangle=\frac{1}{\sqrt{2}}  (\left |010\right \rangle+\left |101\right \rangle),\left |\Psi_6  \right \rangle=\frac{1}{\sqrt{2}}  (\left |010\right \rangle-\left |101\right \rangle)\\
\left |\Psi_7  \right \rangle=\frac{1}{\sqrt{2}}  (\left |011\right \rangle+\left |100\right \rangle),\left |\Psi_8  \right \rangle=\frac{1}{\sqrt{2}}  (\left |011\right \rangle-\left |100\right \rangle)
    \end{split}
\end{equation}

The n-GHZ state is expressed as follows
\begin{equation}
\begin{split}
\left\{\begin{matrix}
 \left |\Psi_1 \right \rangle&=\frac{1}{\sqrt{2}}(\overbrace{\left | 00\dots 0\right \rangle}^{n}+\overbrace{\left | 11\dots 1\right \rangle}^{n})  \\
 \left |\Psi_2 \right \rangle&=\frac{1}{\sqrt{2}}(\overbrace{\left | 00\dots 0\right \rangle}^{n}-\overbrace{\left | 11\dots 1\right \rangle}^{n})\\
 \left |\Psi_3 \right \rangle&=\frac{1}{\sqrt{2}}(\overbrace{\left | 00\dots 1\right \rangle}^{n}+\overbrace{\left | 11\dots 0\right \rangle}^{n})\\
 \left |\Psi_4 \right \rangle&=\frac{1}{\sqrt{2}}(\overbrace{\left | 00\dots 1\right \rangle}^{n}-\overbrace{\left | 11\dots 0\right \rangle}^{n})\\
 \dots \\
 \left |\Psi_{2^n-1} \right \rangle&=\frac{1}{\sqrt{2}}(\overbrace{\left | 01\dots 1\right \rangle}^{n}+\overbrace{\left | 10\dots 0\right \rangle}^{n})\\
 \left |\Psi_{2^n} \right \rangle&=\frac{1}{\sqrt{2}}(\overbrace{\left | 01\dots 1\right \rangle}^{n}-\overbrace{\left | 10\dots 0\right \rangle}^{n})\\
\end{matrix}\right.        
\end{split}
\end{equation}

\subsection{Fidelity}
Fidelity is an important metric for assessing the accuracy of quantum computing results which is influenced by noise, it measures how similar two quantum states are to each other. Before introducing fidelity, we introduce the concept of the density operator. For an arbitrary quantum state $\left | \phi  \right \rangle$ with the following definition of the density operator:
\begin{equation}
\rho =\sum_{i=1}^{N} p_{i} \left | \varphi _{i}   \right \rangle \left\langle\varphi _{i} \right|   \label{eq2}
\end{equation}
where ${\left \{ N|N\in \mathit{N}_+,N>2 \right \} }$, $p_i$ is the probability that a quantum state $\left | \varphi _{i}   \right \rangle $ exists in $\left | \phi \right \rangle$, and $\sum_{i=1}^{N}p_{i} =1$.

Then the fidelity F can be defined as
\begin{equation}
F=\sqrt{\left\langle\psi\right| \rho \left | \psi  \right \rangle } \label{eq3}
\end{equation}

where $\rho$ is the density operator of the resulting quantum state after the measurement and $\left | \psi  \right \rangle$ is the quantum state that should be obtained from the measurement in the ideal case. If the fidelity is closer to 1, it means that this measurement is closer to the ideal correct measurement.

\subsection{Super-dense Coding Protocol Based on GHZ state}
\begin{figure}[t]
\centerline{\includegraphics[width=0.8\linewidth]{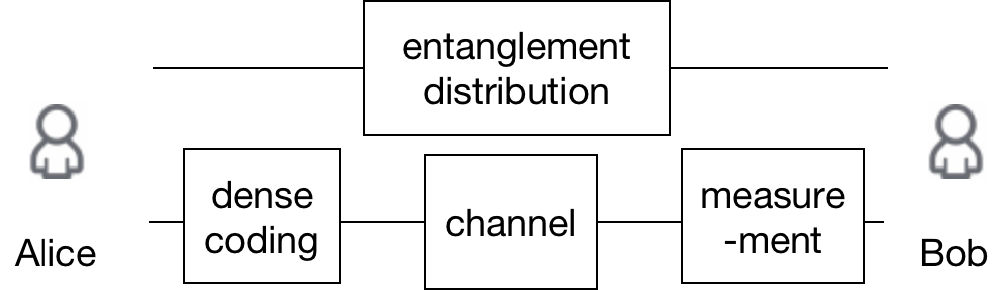}}
\caption{The process of super-dense coding communication.}
\label{fig1}
\end{figure}

The communication process of super-dense coding is shown in Fig.~\ref{fig1}. Assume that Alice and Bob share a pair of GHZ states in advance via entanglement distribution $\left | \Psi^{+}    \right \rangle =\frac{1}{\sqrt{2} } (\left | 000  \right \rangle_{BAA} +\left | 111  \right \rangle_{BAA}  )$, Bob holds the 1st qubit, Alice holds the 2nd and 3rd qubits, and the following is the super-dense coding protocol based on the GHZ state\cite{cereceda2001quantum}\cite{gorbachev2000teleportation}.
\paragraph{Step 1}According to a prior mutually agreed protocol, Alice combines the quantum gates $\left \{ I,\sigma _{x},i\sigma _{y},\sigma _{z} \right \} $ into a unitary operator \textbf{U} as the encoding of the classical bits of information to be transmitted, acting on the 2 held qubits and transmitting them to Bob via the quantum channel.
\paragraph{Step 2}Bob jointly measures the qubits transmitted by Alice with the qubits he holds, obtains Alice's encoded information, and translates it into classical information.

The specific coding protocols are shown in Table ~\ref{tab1}.
Traditional communication methods can only encode one classical message into one binary bit, while quantum-dense coding can encode multiple classical bits into one qubit, greatly improving the efficiency and speed of information transmission.

\begin{table}[h]
\caption{Super-dense Coding Protocol Based on GHZ State}
\begin{center}
\begin{tabular}{|c|c|c|}
\hline
\begin{tabular}[c]{@{}c@{}}Classical\\ Information\end{tabular} & \begin{tabular}[c]{@{}c@{}}Unitary\\ Operator \textbf{U}\end{tabular} & \begin{tabular}[c]{@{}c@{}}Received\\ Entanglement State\end{tabular}                          \\\hline
000                                                             & \textbf{I$\otimes$I}                                                  & $\frac{1}{\sqrt{2} } (\left | 000  \right \rangle _{BAA} +\left | 111  \right \rangle _{BAA})$ \\\hline
001                                                             & \textbf{$\sigma_{z}\otimes$I}                                                  & $\frac{1}{\sqrt{2} } (\left | 000  \right \rangle _{BAA} -\left | 111  \right \rangle _{BAA})$ \\\hline
010                                                             & \textbf{I$\otimes\sigma_{x}$}                                                  & $\frac{1}{\sqrt{2} } (\left | 001  \right \rangle _{BAA} +\left | 110  \right \rangle _{BAA})$ \\\hline
011                                                             & \textbf{$\sigma _{z}\otimes\sigma _{x}$}                                                  & $\frac{1}{\sqrt{2} } (\left | 001  \right \rangle _{BAA} -\left | 110  \right \rangle _{BAA})$ \\\hline
100                                                             & \textbf{$\sigma _{x}\otimes$I}                                                  & $\frac{1}{\sqrt{2} } (\left | 010  \right \rangle _{BAA} +\left | 101  \right \rangle _{BAA})$ \\\hline
101                                                             & -i\textbf{$\sigma _{y}\otimes$I}                                                & $\frac{1}{\sqrt{2} } (\left | 010  \right \rangle _{BAA} -\left | 101  \right \rangle _{BAA})$ \\\hline
110                                                             & \textbf{$\sigma_{x}\otimes\sigma_{x}$}                                                  & $\frac{1}{\sqrt{2} } (\left | 011  \right \rangle _{BAA} +\left | 100  \right \rangle _{BAA})$ \\\hline
111                                                             & -i\textbf{$\sigma_{y}\otimes\sigma_{x}$}                                                & $\frac{1}{\sqrt{2} } (\left | 011  \right \rangle _{BAA} -\left | 101  \right \rangle _{BAA})$\\ \hline
\end{tabular}
\label{tab1}
\end{center}
\end{table}

\subsection{Entanglement Purification Based on GHZ State}
Bennett et al. \cite{bennett1996purification} first proposed entanglement purification in 1997, and developed a standard entanglement purification protocol. Later, Deutsch et al. improved entanglement purification\cite{deutsch1996quantum} by using quantum privacy amplification to transform phase-flip noise into bit-flip noise for subsequent processing to improve entanglement purification performance. The entanglement purification method in this paper is developed mainly based on the work of Deutsch. 

\begin{figure}[t]
\centerline{\includegraphics[width=0.9\linewidth]{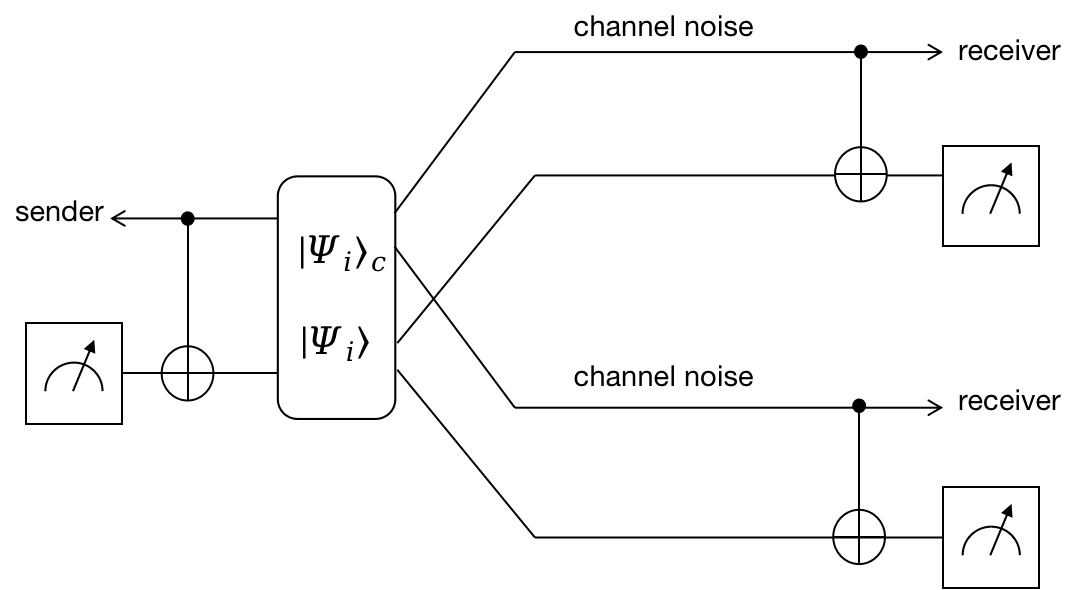}}
\caption{The process of entanglement purification based on GHZ state.}
\label{fig2}
\end{figure}

A simplified circuit diagram of an entanglement purification process based on GHZ states is shown in Fig.~\ref{fig2}. Assuming that the sender holds two pairs of entangled GHZ states, the sender sends half of the two pairs of entangled states to the receiver via entanglement distribution. Next, each side performs a CNOT gate operation on the qubits it holds and measures the target group qubits. If the qubits of the target group are the same, the qubits of the control group are kept and the qubits of the target group are discarded. If the qubits of the target group are different, then both groups are discarded.

\begin{figure}[t]
\centerline{\includegraphics[width=0.8\linewidth]{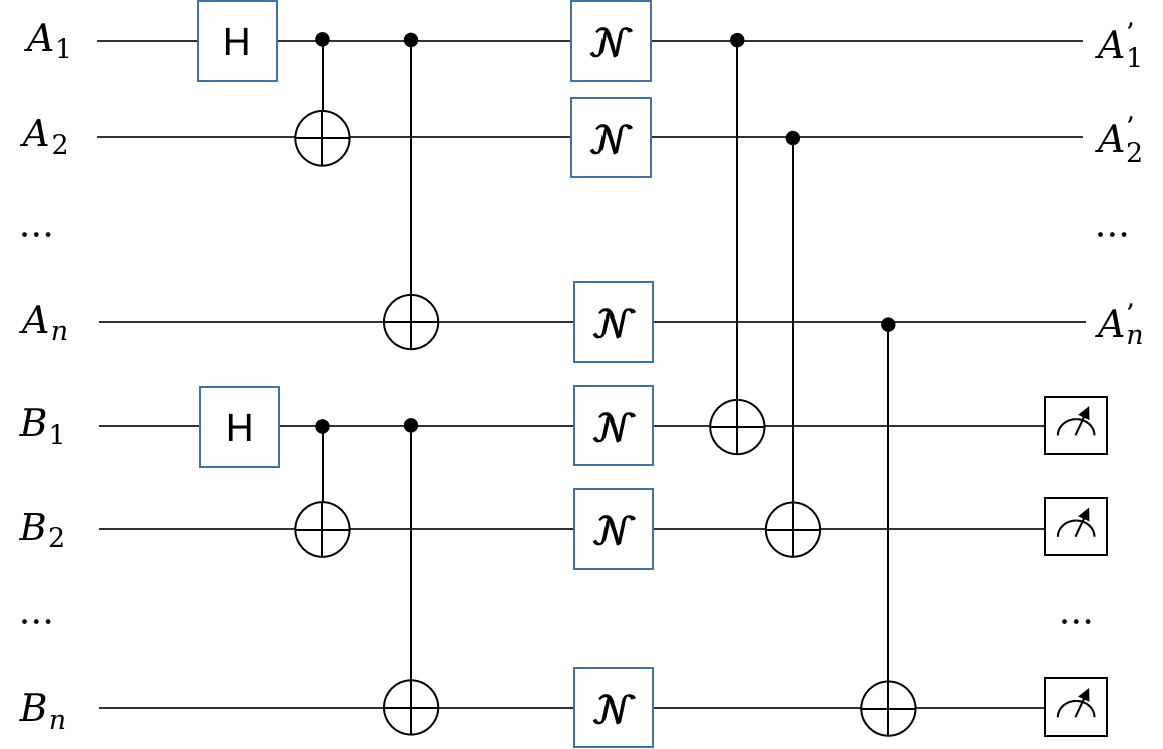}}
\caption{A n-dimensional entanglement purification circuit.}
\label{fig3}
\end{figure}

It is worth mentioning that since the real channel contains not only bit-flip noise but also phase-flip noise, the quantum channel identification technique in \cite{beer2020training} can be used to identify the type of noise first, and for phase-flip noise, it can be converted to bit-flip noise using H-gate before and after transmission.

An entanglement purification circuit dealing with n-dimensional entangled states is shown in Fig.~\ref{fig3}. Where $A_i$ is the qubits of the control group quantum states and $B_j$ is the qubits of the target group quantum states. The decision to retain the control group A is made by measuring the target group B. It is a waste of qubit resources and does not provide a satisfactory fidelity improvement if only one entanglement purification is carried out. Therefore, there is a need to find a way to improve the efficiency of entanglement purification while saving qubit resources.

\subsection{Quantum Neural Network}
The basic neuron of the QNN is the quantum perceptron, which can be regarded as a unitary operator that can process the input quantum state to obtain the output quantum state. In the process of training the QNN, the network iterates continuously and the final model obtained is also a unitary operator, using which the input data to be processed can be computed directly.

\begin{figure}[t]
\centerline{\includegraphics[width=0.75\linewidth]{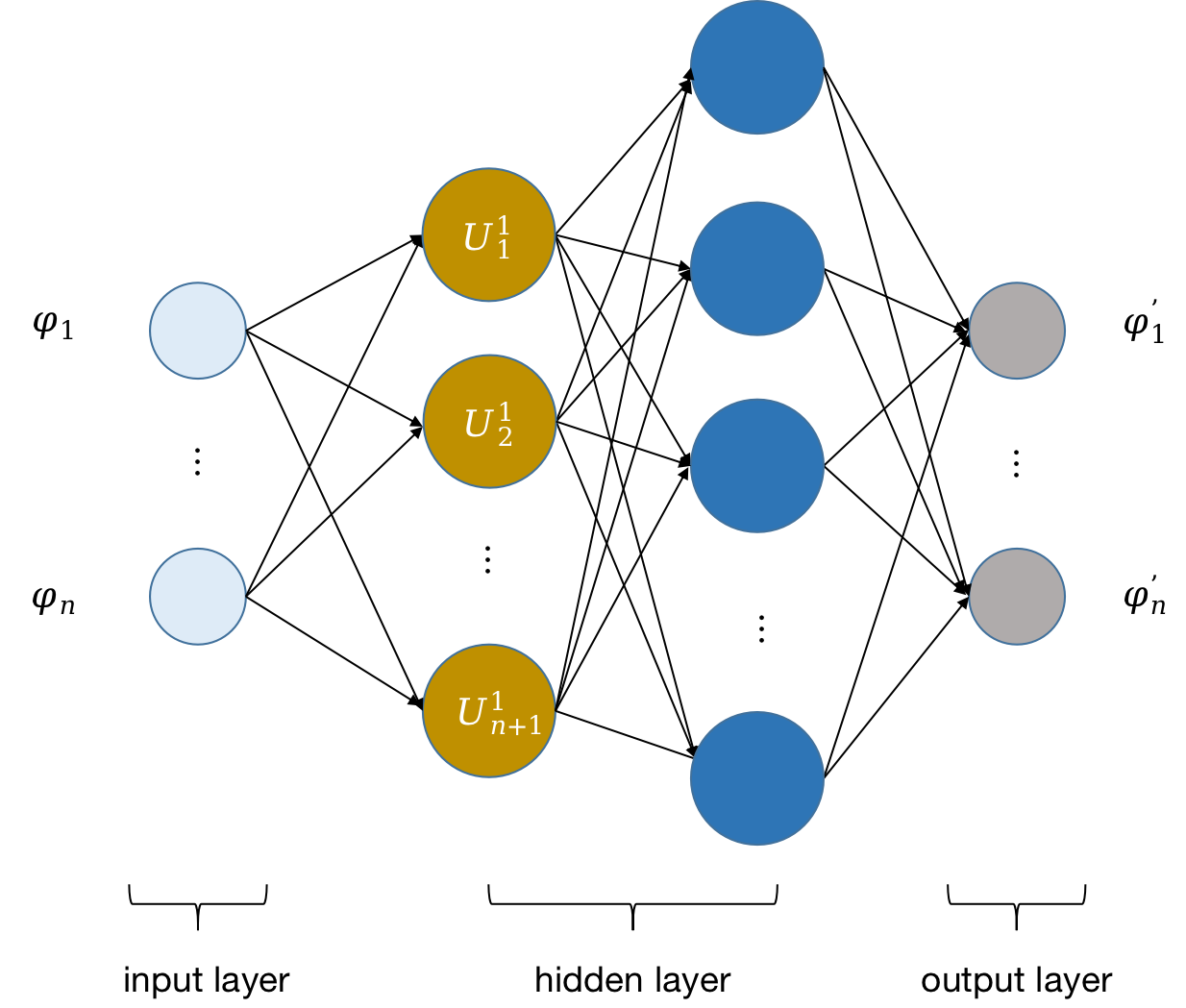}}
\caption{The quantum neural network structure.}
\label{fig4}
\end{figure}

The general structural diagram of the quantum neural network QNN used is shown in Fig.~\ref{fig4}.

As can be seen from Fig.~\ref{fig4}, the QNN network structure is exactly adapted to the structure of the purified circuit. We can use the known ideal entanglement distribution results to improve the performance of the purification circuit by using the QNN to train a unitary operator that can express the effects of noise.

As shown in Fig.~\ref{fig4}, both the input and output layers are of dimension n. The output of the n-dimensional purification circuit will be used as the training set input to the QNN. There are L hidden layers in between, where the unitary operator $U^l$ of each layer is obtained by multiplying multiple quantum perceptrons in sequence, i.e.
\begin{equation}
U^{l}=U_{n+1}^{l}\dots U_{2}^{l}U_{1}^{l}
\end{equation}

Assuming that the output quantum state of the n-dimensional purification circuit is $\rho _{\Psi }$, and after L-layer hidden layer iterations, the output quantum state $\rho _{\Psi }^{'} $ is obtained, we have
\begin{equation}
    \rho _{\Psi }^{'}=tr_{in,hid}(U(\rho _{\Psi }\otimes \left | 0\dots 0  \right \rangle_{ hid,out}\left\langle0\dots 0\right|   )U^{\dagger } )  
\end{equation}
where $U$ is the unitary operator circuit obtained from the training.

The training set of the QNN is assumed to be $\left \{ (\left | \Psi _{1}^{in}   \right \rangle,\left | \Psi _{1}^{out}   \right \rangle),(\left | \Psi _{2}^{in}   \right \rangle,\left | \Psi _{2}^{out}   \right \rangle),\dots,(\left | \Psi _{N}^{in}   \right \rangle,\left | \Psi _{N}^{out}   \right \rangle) \right \} $, QNN uses fidelity as a loss function to measure how well the network is trained, i.e. with training input $\left | \Psi _{x}^{in}   \right \rangle(x=1,2,..., N)$, the output of the quantum neural network $\rho_{x}^{out}$ is similar to $\left | \Psi _{x}^{in}   \right \rangle$. The loss function C is calculated as follows:
\begin{equation}
   C=\frac{1}{N}\sum_{x=1}^{N}\left\langle\Psi _{x}^{out} \right| \rho _{x}^{out}\left | \Psi _{x}^{out}   \right \rangle    
\end{equation}

\subsection{Quantum Channel Capacity}
For a general quantum channel, assume any quantum state $\left | \Psi\right \rangle $ in Hilbert space with density operator $\rho$ and eigenvalue $\lambda_i$, where $i=1,2,..., N$ and $N$ is the number of dimensions in Hilbert space. Corresponding to the average mutual information in the classical channel is the quantum coherent information $I_{coh}(\rho, N) $. The classical capacity in a quantum channel is defined by Quantum mutual information, which measures the ability of the quantum channel to transmit classical information. Quantum capacity is defined by Quantum coherent information, which measures the ability of a quantum channel to transmit quantum information.

Assuming that the input to the quantum channel is $X_{in}=\left \{ \left |\Psi _{1} \right \rangle,\left |\Psi _{2} \right \rangle,\dots \left |\Psi _{N} \right \rangle  \right \} $, when the input is $\left |\Psi _{i} \right \rangle$, the output quantum state $\rho_i$ is obtained with probability $\pi_i$.
The classical average mutual information $I_{c}(\rho ,N)$ of the quantum channel can be derived from the Holevo bound\cite{imre2012advanced}
\begin{equation}
   I_{c}(\rho ,N)=S(\pi _{i}\rho _{i})-\sum_{i=1}^{N}\pi _{i}S(\rho _{i})   
\end{equation}
where $S(\cdot)$ is the von Neumann entropy\cite{nielsen2002quantum}
\begin{equation}
  S(\rho)=-Tr(\rho\log{\rho})=-\sum_{i=1}^{N}\lambda_{i}\log\lambda_{i}  
\end{equation}

The classical capacity of a quantum channel is defined as
\begin{equation}
C=\max_{\pi _{i},\rho _{i}  } I_{c}(\rho,N)
\end{equation}
The quantum coherent information of a quantum channel\cite{grishanin2000coherent} is defined as
\begin{equation}
I_{coh}=S(\pi _i\rho _i)-S_e
\end{equation}
where $S_e$ can be seen as the amount of information exchanged with the environment as the system interacts with it \cite{9998300}\cite{schumacher1996sending}\cite{lloyd1997capacity}, and it characterizes the amount of quantum noise during the evolution of the system. the expression for $S_e$ is as follows
\begin{equation}
    S_e=S(\sum_{i,j}\mathcal{N}(\left | \psi_{i}  \right \rangle\left\langle\psi_{j} \right| )\otimes\left | \overline{\psi_{i}}  \right \rangle\left\langle\overline{\psi_{j}} \right|)
\end{equation}
where $\mathcal{N}$ denotes the channel noise, $\left | \psi_{i} \right \rangle=\pi_{i}^{\frac{1}{4} }\left |\Psi _{i}  \right \rangle $ is the eigenvector of the input quantum state density matrix, and $\left | \overline{\psi_{i}}  \right \rangle$ is its complex conjugate vector.
This gives the quantum capacity $C_Q$ of the quantum channel as
\begin{equation}
    C_Q=\max_{\pi _i,\rho _i}I_{coh} \label{eq18}
\end{equation}

\section{Super-dense Coding Protocol Based on n-GHZ state}\label{Section3}
To transmit more quantum states, this paper proposes an ultra-dense coding protocol based on n-GHZ states. Assuming that Alice needs to transmit n bits of classical information, Alice transmits one qubit of the n-GHZ state to Bob via entanglement distribution, and encodes the classical information $X=x_{n-1} x_{n-2}\dots x_{0}$ into the qubits Alice held using the unitary operator $U_{sdc}$ in \eqref{eq77}.

\begin{equation}
\begin{split}
\label{eq77}
    U_{sdc}  = &\left \{(1-x_0)[(1-x_{n-1})I+x_{n-1}\sigma_x ]+x_0[(1-x_{n-1})\sigma_z \right.
\\ &\left.-ix_{n-1}\sigma_y] \right \}\otimes[(1-y_{k} )I+y_k\sigma_x]^{\otimes(n-2)}\\ 
\end{split}
\end{equation}
\begin{equation}
    \left \lfloor \frac{X}{2}  \right \rfloor = y_{n-2}y_{n-3}\dots y_0  
\end{equation}

where $\left \lfloor \cdot  \right \rfloor$ is the round-down symbol, $k=n-3,n-2,...,0$. $U_{sdc}$ is a $n-1$ dimensional unitary operator.

Alice transmits the encoded qubits through the quantum channel to Bob, who then makes a joint measurement of the qubits he holds with the qubits transmitted by Alice to restore the classical information Alice wants to transmit.

\section{Experimental simulation and analysis}\label{Section4}
In this paper, the input of the training set of the QNN network is the shared quantum state after entanglement distribution in the actual noise environment, and the output is the ideal shared quantum state after entanglement distribution. Depolarization noise which is unitary and amplitude damping noise which is non-unitary are chosen as the channel noise to train the network respectively, and the results are analysed and compared according to the results.

\begin{figure}[htbp]
\centerline{\includegraphics[width=1\linewidth]{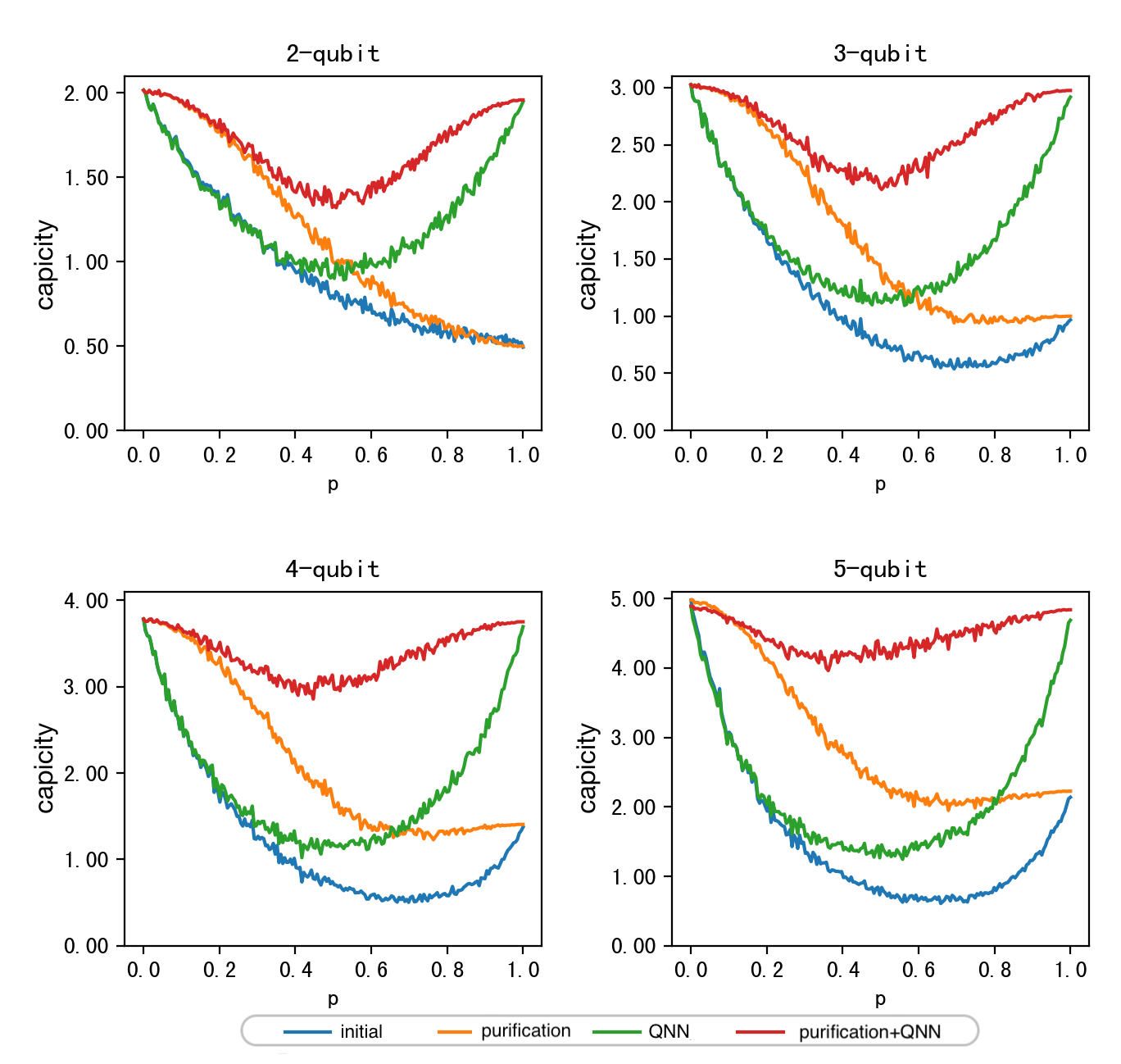}}
\caption{Channel capacity of super-dense coding under amplitude damping noise.}
\label{fig5}
\end{figure}

\begin{figure}[htbp]
\centerline{\includegraphics[width=1\linewidth]{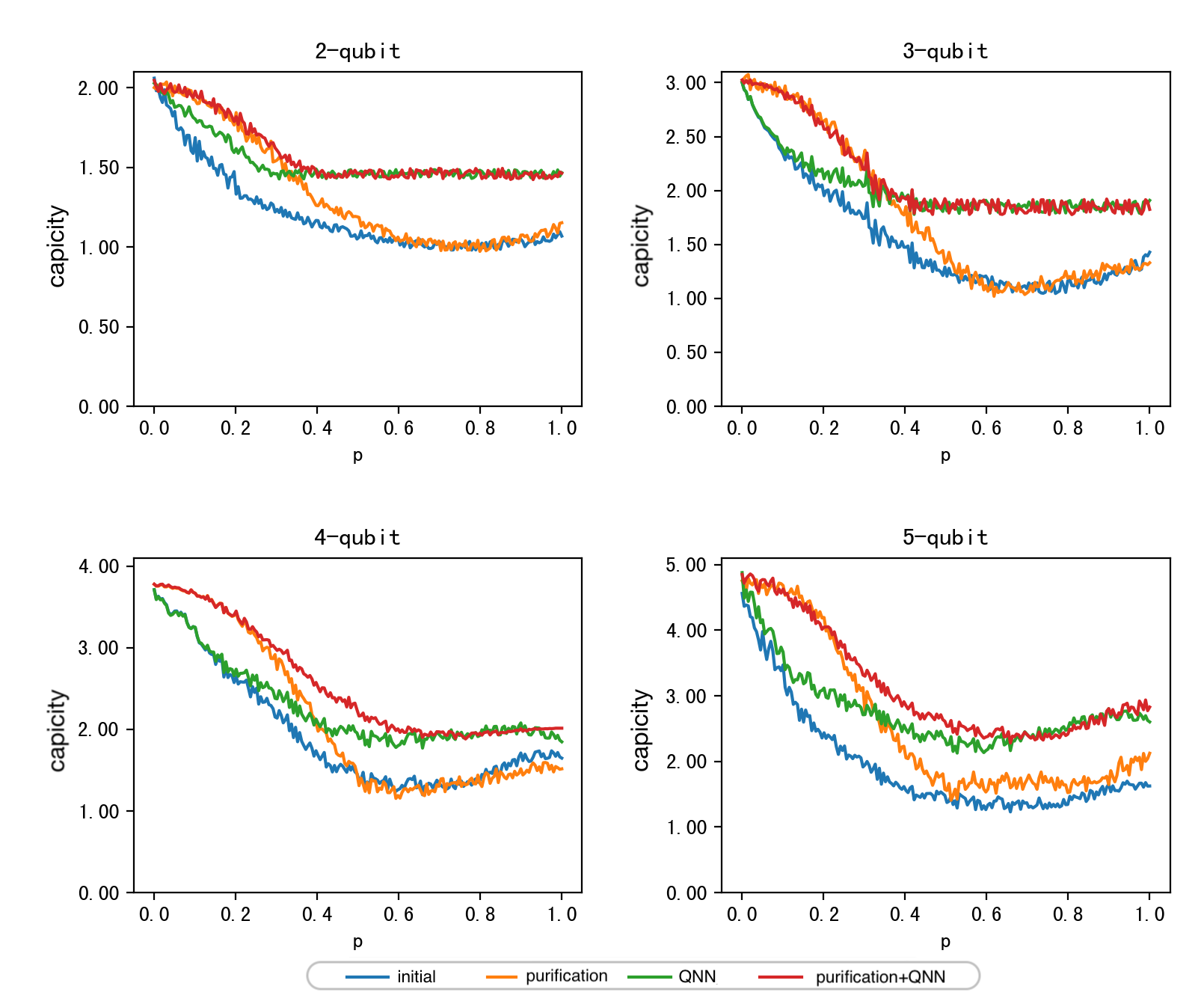}}
\caption{Channel capacity of super-dense coding under depolarization noise.}
\label{fig6}
\end{figure}

The input to the training set is generated by drawing the quantum circuit in Fig.~\ref{fig3} by the Cirq platform provided by Google, which can simulate a realistic quantum noise environment. The channel capacity simulation results for super-dense coding with amplitude damping noise and depolarization noise are shown in Fig.~\ref{fig5} and Fig.~\ref{fig6}, where $p$ is the noise error probability.

According to the simulation results, under both amplitude damping noise and depolarization noise conditions, the channel capacity decreases and then increases slightly with the increase of noise probability $p$. However, after $p>0.2$, the channel capacity is less than $\frac{1}{2} $ of the ideal channel capacity value. In contrast, when the quantum states are purified after the entanglement distribution and then transmitted by super-dense coding, the channel capacity improves with the decrease of $p$ and slows down, but the channel capacity value is still low at higher $p$ values. If the shared quantum state after entanglement distribution is processed using the trained QNN network operator $U_Q$, the channel capacity remains close to the ideal channel capacity value for noise probability $p$ close to 1 under amplitude damping noise conditions, which is a great improvement to the situation where the channel capacity is degraded by noise, but in the interval of $0.2\le p\le0.8$, the channel capacity still drops to less than half of the ideal channel capacity value. For depolarization noise, the QNN effect on channel capacity enhancement is reflected in a different region from that of purification, where purification has a more significant effect on channel capacity enhancement in the region with lower $p$ values, whereas the QNN effect on channel capacity enhancement is reflected in the region with higher $p$ values. If the purified quantum states are processed using $U_Q$, combining purification and QNN, the enhancement effect on channel capacity is more significant, and the curves show the superposition of the enhancement effect of both purification and QNN.

From Fig.~\ref{fig5} and Fig.~\ref{fig6}, it can be seen that this scheme is not as effective in improving the capacity of the super-dense coding channel under depolarization noise as under amplitude damping noise, presumably because depolarization noise, as unitary noise, will affect the fidelity of the shared quantum state after the entanglement distribution, but will not disentangle the shared quantum state, whereas the essence of both entanglement purification and QNN is to further improve the fidelity of the shared quantum state by maintaining the entanglement of the shared quantum state as much as possible, thus the effect of depolarization noise on channel capacity is not only due to fidelity but also for other reasons.

\begin{figure}[t]
\centerline{\includegraphics[width=1\linewidth]{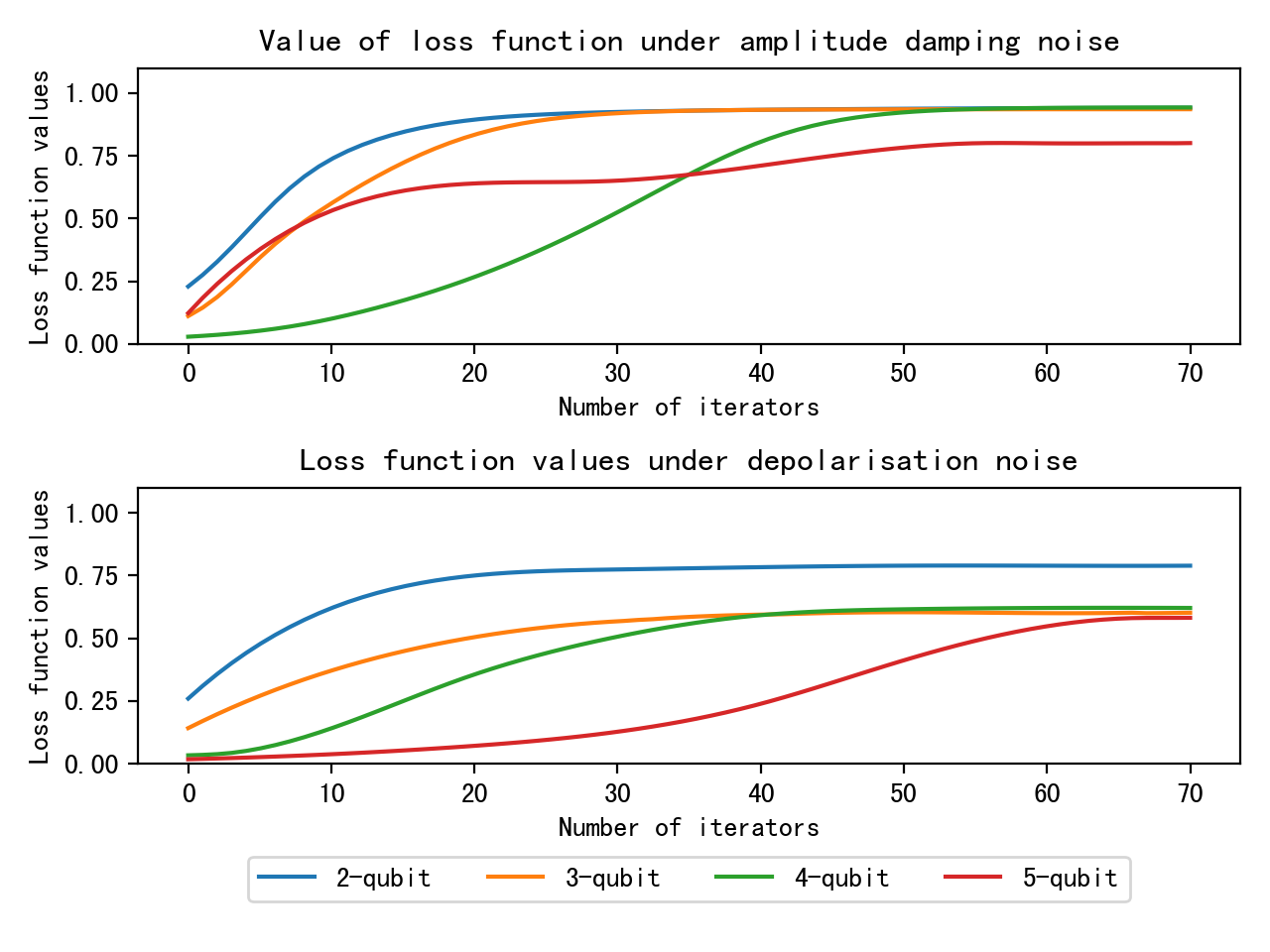}}
\caption{Channel capacity of super-dense coding under depolarization noise.}
\label{fig7}
\end{figure}

The convergence of the loss function of the QNN network training for different numbers of input qubits is shown in Fig.~\ref{fig7},
and the training time of the QNN network and the convergence values of the loss function under different noises are shown in Table ~\ref{tab3}. Where $C_a$ is the convergence value of the loss function under amplitude damping noise and $C_d$ is the convergence value of the loss function under depolarization noise.
\begin{table}[t]
\caption{Super-dense Coding Protocol Based on GHZ State}
\begin{center}
\begin{tabular}{|c|c|c|c|}
\hline
\begin{tabular}[c]{@{}c@{}}Input\\ Dimension\end{tabular} & \begin{tabular}[c]{@{}c@{}}Training\\ Time\end{tabular} & \begin{tabular}[c]{@{}c@{}}Convergence \\ Values $C_a$\end{tabular} & \begin{tabular}[c]{@{}c@{}}Convergence \\ Values $C_d$\end{tabular} \\ \hline
2                                                         & 25.829s                                                 & 0.9418                                                              & 0.7902                                                              \\ \hline
3                                                         & 47.874s                                                 & 0.9359                                                              & 0.6041                                                              \\ \hline
4                                                         & 188.588s                                                & 0.9428                                                              & 0.6218                                                              \\ \hline
5                                                         & 295.908s                                                & 0.8010                                                              & 0.5820                                                              \\ \hline
\end{tabular}
\label{tab3}
\end{center}
\end{table}
It can be seen that the convergence value of the loss function of the QNN network drops sharply when the input qubit count reaches 5 under both noises, and the network is not trained well. As the number of input qubits increases, the convergence of the loss function gradually becomes slower, and when the number of qubits is greater than 5, the loss function tends to fail to converge because of the gradient explosion. The depolarization noise relative to the amplitude damping noise results in a lower convergence value of the final loss function, which is even less effective in training.

\section{Conclusions}\label{Section5}
In this paper, we derive a super-dense coding protocol based on the n-GHZ state based on the dense protocol of the GHZ state. The n-GHZ state super-dense coding protocol applies to the transmission of n+1 bit classical bit information with n-bit qubits in a two-party dense coding communication scenario. It is also focused on using purification and QNN to improve the fidelity of the shared quantum state after entanglement distribution. Based on the channel capacity calculation method of super-dense coding, it is reasoned that the improvement of the fidelity of the shared quantum state after entanglement distribution can bring a certain improvement to the channel capacity of super-dense coding, and this conclusion is proved experimentally. The combination of purification and QNN has a superposition effect on the improvement of the fidelity of the entanglement distribution and therefore can increase the channel capacity of super-dense coding to a certain extent under both unitary and non-unitary noise.

However, there are still some problems in this paper. Firstly, the gradient explosion in QNN network training at input qubit numbers greater than 5 has not yet found a way to be solved, and further research and improvement of the network structure of QNN may be needed to adapt to higher dimensional training set inputs. Secondly, the effect of entanglement purification fidelity enhancement on the capacity of superdense coded channels cannot be verified in a real super-dense coding communication environment due to the lack of practical measurement devices.

\bibliographystyle{IEEEtran}
\bibliography{IEEEabrv,mylib}

\begin{thebibliography}{10}
\providecommand{\url}[1]{#1}
\csname url@samestyle\endcsname
\providecommand{\newblock}{\relax}
\providecommand{\bibinfo}[2]{#2}
\providecommand{\BIBentrySTDinterwordspacing}{\spaceskip=0pt\relax}
\providecommand{\BIBentryALTinterwordstretchfactor}{4}
\providecommand{\BIBentryALTinterwordspacing}{\spaceskip=\fontdimen2\font plus
\BIBentryALTinterwordstretchfactor\fontdimen3\font minus \fontdimen4\font\relax}
\providecommand{\BIBforeignlanguage}[2]{{%
\expandafter\ifx\csname l@#1\endcsname\relax
\typeout{** WARNING: IEEEtran.bst: No hyphenation pattern has been}%
\typeout{** loaded for the language `#1'. Using the pattern for}%
\typeout{** the default language instead.}%
\else
\language=\csname l@#1\endcsname
\fi
#2}}
\providecommand{\BIBdecl}{\relax}
\BIBdecl

\bibitem{vathsan2015introduction}
R.~Vathsan, \emph{Introduction to quantum physics and information processing}.\hskip 1em plus 0.5em minus 0.4em\relax CRC Press, 2015.

\bibitem{9797967}
Y.~Wang and X.~Chen, ``Joint modulation of 3-ppm and quantum squeezed states in communication systems,'' in \emph{IEEE INFOCOM 2022 - IEEE Conference on Computer Communications Workshops (INFOCOM WKSHPS)}, 2022, pp. 1--5.

\bibitem{bennett1992communication}
C.~H. Bennett and S.~J. Wiesner, ``Communication via one-and two-particle operators on einstein-podolsky-rosen states,'' \emph{Physical review letters}, vol.~69, no.~20, p. 2881, 1992.

\bibitem{mattle1996dense}
K.~Mattle, H.~Weinfurter, P.~G. Kwiat, and A.~Zeilinger, ``Dense coding in experimental quantum communication,'' \emph{Physical Review Letters}, vol.~76, no.~25, p. 4656, 1996.

\bibitem{long2002theoretically}
G.-L. Long and X.-S. Liu, ``Theoretically efficient high-capacity quantum-key-distribution scheme,'' \emph{Physical Review A}, vol.~65, no.~3, p. 032302, 2002.

\bibitem{bennett1996concentrating}
C.~H. Bennett, H.~J. Bernstein, S.~Popescu, and B.~Schumacher, ``Concentrating partial entanglement by local operations,'' \emph{Physical Review A}, vol.~53, no.~4, p. 2046, 1996.

\bibitem{lo2001concentrating}
H.-K. Lo and S.~Popescu, ``Concentrating entanglement by local actions: Beyond mean values,'' \emph{Physical Review A}, vol.~63, no.~2, p. 022301, 2001.

\bibitem{bennett1996purification}
C.~H. Bennett, G.~Brassard, S.~Popescu, B.~Schumacher, J.~A. Smolin, and W.~K. Wootters, ``Purification of noisy entanglement and faithful teleportation via noisy channels,'' \emph{Physical review letters}, vol.~76, no.~5, p. 722, 1996.

\bibitem{tacchino2019artificial}
F.~Tacchino, C.~Macchiavello, D.~Gerace, and D.~Bajoni, ``An artificial neuron implemented on an actual quantum processor,'' \emph{npj Quantum Information}, vol.~5, no.~1, p.~26, 2019.

\bibitem{cereceda2001quantum}
J.~L. Cereceda, ``Quantum dense coding using three qubits,'' \emph{arXiv preprint quant-ph/0105096}, 2001.

\bibitem{gorbachev2000teleportation}
V.~Gorbachev, A.~Trubilko, A.~Zhiliba, and E.~Yakovleva, ``Teleportation of entangled states and dense coding using a multiparticle quantum channel,'' \emph{arXiv preprint quant-ph/0011124}, 2000.

\bibitem{deutsch1996quantum}
D.~Deutsch, A.~Ekert, R.~Jozsa, C.~Macchiavello, S.~Popescu, and A.~Sanpera, ``Quantum privacy amplification and the security of quantum cryptography over noisy channels,'' \emph{Physical review letters}, vol.~77, no.~13, p. 2818, 1996.

\bibitem{beer2020training}
K.~Beer, D.~Bondarenko, T.~Farrelly, T.~J. Osborne, R.~Salzmann, D.~Scheiermann, and R.~Wolf, ``Training deep quantum neural networks,'' \emph{Nature communications}, vol.~11, no.~1, p. 808, 2020.

\bibitem{imre2012advanced}
S.~Imre and L.~Gyongyosi, \emph{Advanced quantum communications: an engineering approach}.\hskip 1em plus 0.5em minus 0.4em\relax John Wiley \& Sons, 2012.

\bibitem{nielsen2002quantum}
M.~A. Nielsen and I.~Chuang, ``Quantum computation and quantum information,'' 2002.

\bibitem{grishanin2000coherent}
B.~Grishanin and V.~Zadkov, ``Coherent-information analysis of quantum channels in simple quantum systems,'' \emph{Physical Review A}, vol.~62, no.~3, p. 032303, 2000.

\bibitem{9998300}
Y.~Wang, R.~Zhang, B.~Lu, and X.~Chen, ``Optimal detection with squeezed state for quantum communication in thermal noise channel,'' in \emph{2022 4th IEEE Middle East and North Africa COMMunications Conference (MENACOMM)}, 2022, pp. 43--48.

\bibitem{schumacher1996sending}
B.~Schumacher, ``Sending entanglement through noisy quantum channels,'' \emph{Physical Review A}, vol.~54, no.~4, p. 2614, 1996.

\bibitem{lloyd1997capacity}
S.~Lloyd, ``Capacity of the noisy quantum channel,'' \emph{Physical Review A}, vol.~55, no.~3, p. 1613, 1997.

\end{thebibliography}

\vspace{12pt}
\end{document}